\documentclass[conference]{IEEEtran}
\IEEEoverridecommandlockouts
\usepackage{cite}
\usepackage{amsmath,amssymb,amsfonts}
\usepackage{graphicx}
\usepackage{textcomp}
\usepackage{xcolor}

\usepackage{orcidlink}
\usepackage{hyperref}
\usepackage{makecell}
\usepackage{tabularx}
\usepackage{acronym}
\usepackage{algorithm}
\usepackage{algpseudocode}
\usepackage{booktabs}

\acrodef{ML}{Machine Learning}
\acrodef{FL}{Federated Learning}
\acrodef{SMC}{Secure Multi-party Computation}
\acrodef{ZK}{Zero-Knowledge}

\def\BibTeX{{\rm B\kern-.05em{\sc i\kern-.025em b}\kern-.08em
    T\kern-.1667em\lower.7ex\hbox{E}\kern-.125emX}}
\begin{document}

\title{Verifiability and Privacy in Federated Learning through Context-Hiding Multi-Key Homomorphic Authenticators}


\author{\IEEEauthorblockN{Simone Bottoni\orcidlink{0000-0003-3758-5666}}
\IEEEauthorblockA{
\textit{University of Insubria}\\
Varese, Italy \\
simone.bottoni@uninsubria.it}
\and
\IEEEauthorblockN{Giulio Zizzo\orcidlink{0009-0004-5750-5744}}
\IEEEauthorblockA{
\textit{IBM Research Europe}\\
Dublin, Ireland \\
giulio.zizzo2@ibm.com}
\and
\IEEEauthorblockN{Stefano Braghin\orcidlink{0000-0001-5519-1674}}
\IEEEauthorblockA{
\textit{IBM Research Europe}\\
Dublin, Ireland \\
stefanob@ie.ibm.com}
\and
\IEEEauthorblockN{Alberto Trombetta\orcidlink{0000-0002-2567-9297}}
\IEEEauthorblockA{
\textit{University of Insubria}\\
Varese, Italy \\
alberto.trombetta@uninsubria.it}
}

\maketitle

\begin{abstract}
Federated Learning has rapidly expanded from its original inception to now have a large body of research, several frameworks, and sold in a variety of commercial offerings. Thus, its security and robustness is of significant importance. There are many algorithms that provide robustness in the case of malicious clients. However, the aggregator itself may behave maliciously, for example, by biasing the model or tampering with the weights to weaken the model’s privacy.
In this work, we introduce a verifiable federated learning protocol that enables clients to verify the correctness of the aggregator’s computation without compromising the confidentiality of their updates. Our protocol uses a standard secure aggregation technique to protect individual model updates with a linearly homomorphic authenticator scheme that enables efficient, privacy-preserving verification of the aggregated result. Our construction ensures that clients can detect manipulation by the aggregator while maintaining low computational overhead. We demonstrate that our approach scales to large models, enabling verification over large neural networks with millions of parameters.
\end{abstract}


\section{Introduction}
\label{sec:introduction}

Current advancements in Generative AI are made possible because of large public datasets or in-house dataset curation by organizations working on such models.
In several domains, including crucial sectors like finance and healthcare, it is not possible to accumulate such an overwhelming amount of data to train effective and accurate models.
There are several reasons for this, most of which rely on the inherent confidentiality of the data needed to train such models.

Hence, alternative solutions have been explored to train AI models without directly sharing datasets.
These techniques are generally grouped under the Trustworthy AI topic~\cite{zheng2024overview}, which includes methods and techniques for training AI models in a secure, privacy-preserving, fair, and verifiable fashion.

In this context, \ac{FL} is a well-established paradigm that allows entities, such as government organizations, companies, and hospitals, to collaboratively train a \ac{ML} model in a distributed fashion without sharing their private plaintext data.
With the help of \ac{FL}, such entities, known as clients, collect and hold the data, train a local \ac{FL} model, and share the updated model's parameters with each other or with a third-party entity called the aggregator. The aggregator then combines these individual contributions into a new global model.
This process is repeated over several rounds to iteratively refine the global model.
By the end of the training process, the clients will have a shared \ac{FL} model trained on the combined data from all entities.

Although \ac{FL} is advocated as a privacy-enhancing technology\footnote{\url{https://ico.org.uk/for-organisations/uk-gdpr-guidance-and-resources/data-sharing/privacy-enhancing-technologies/}}, in its default implementation has several possible attacks and security issues. For example, \ac{FL} systems are vulnerable to inference attacks~\cite{DBLP:journals/corr/abs-2012-06337} where an attacker (such as a client or the aggregator) can infer sensitive information about other participants' private data from the \ac{ML} algorithm parameters~\cite{DBLP:conf/sp/MelisSCS19}.

While there is a significant body of research around protecting \ac{FL} systems against malicious clients~\cite{blanchard2017machine, xie2021crfl, zizzo2021certified}, the aggregator remains a critical vulnerability. As the central entity responsible for collecting and aggregating clients' parameters, it can act dishonestly, which enables essentially arbitrary manipulation and tampering.
The aggregator may bias the overall model towards unfair outcomes, disregard contributions by certain clients or user groups, or insert one or more backdoors into the model to control its predictions given targeted inputs~\cite{xie2019dba}.
This is in addition to the privacy concerns posed by a dishonest aggregator, where in addition to potentially launching privacy attacks against specific clients, it can tamper with the weights to make privacy attacks even easier with so-called ``weight trap attacks"~\cite{boenisch2021curious}.

Thus, to combat these problems, verifying the aggregator's behavior in \ac{FL} has started to receive more attention. Existing works have proposed using schemes based on homomorphic hashes~\cite{Fiore2014EfficientlyVC} or Pallier-based encryption~\cite{madi2021secure, bottoni2021privacy}. However, shortfalls remain, either due to prohibitively long running times or lack of privacy protections.

In this work, we propose a novel \ac{FL} protocol that ensures both privacy and verifiability of model aggregation. Our construction enables clients to verify that the aggregator has performed the computation correctly without revealing their private model updates in plaintext. This is achieved by combining a standard masking-based Secure Aggregation scheme, ensuring the confidentiality of individual updates, with a Context-Hiding Identity-Based Multi-Key Linearly Homomorphic Authenticators scheme that guarantees integrity and verifiability of the aggregation process.

Clients authenticate vectors of model parameters using the homomorphic authenticator, enabling the server to aggregate both masked updates and corresponding signatures efficiently. The homomorphic property allows each client to verify the correctness of the aggregated result without compromising clients privacy. We reduce computational and communication overhead by authenticating vectors instead of individual elements, ensuring practicality for large-scale \ac{FL} deployments.

Our main contributions are summarized as follows:
\begin{itemize}
    \item We propose a verifiable \ac{FL} protocol that ensures both the privacy of client updates and the integrity of server aggregation;
    \item We integrate a linearly homomorphic authenticator scheme into a \ac{FL} model to enable efficient and privacy-preserving verification of aggregated results;
    \item We implemented the protocol and we present experiments showing that our construction scales to \ac{FL} models with millions of parameters. The implementation will be released as open source\footnote{Reference to be provided upon paper acceptance}.
\end{itemize}
    
The remainder of the paper is structured as follows. Section~\ref{sec:related} provides a summary of the related works. Section~\ref{sec:preliminaries} offers an overview of the cryptographic primitives we used to ensure the verifiability and privacy of the model updates. Section~\ref{sec:protocol} presents a detailed description of our protocol, outlining the roles of involved parties, their interactions, and an evaluation of its security. Section~\ref{sec:experiments} details the conducted experiments and the results obtained. Finally, Section~\ref{sec:conclusion} concludes our contribution and depicts future research directions.
\section{Related Works}
\label{sec:related}
Federated Learning enables multiple parties to train \ac{ML} models in a distributed way without sharing their private data. Typically, clients send their model updates to a third-party entity known as an aggregator, which combines these updates to create a global model trained on the collective data. However, \ac{FL} is susceptible to inference attacks, as outlined by~\cite{DBLP:journals/corr/abs-2012-06337}, where the aggregator may act maliciously, cheating or tampering with the clients' model weights.

To ensure the privacy of data used in training \ac{FL} models, various techniques have been employed, including \ac{SMC}~\cite{SecureML, Blind_Justice, Sharemind, ABY3}, Differential Privacy\cite{LDP-Fed}, and Homomorphic Encryption~\cite{BatchCrypt}. However, implementing a comprehensive security layer for \ac{FL} models using these techniques individually is challenging. Recent works have attempted to combine these methods~\cite{10.1145/3338501.3357370, HybridAlpha} to enhance privacy. For instance, in~\cite{bottoni2021privacy}, the authors used an \ac{SMC} protocol in combination with Partially Homomorphic Encryption using the Paillier scheme. Despite these advancements, issues such as long execution times, reduced model accuracy, and insufficient privacy protections remain.

Several recent studies have introduced trusted layers composed of various techniques around \ac{FL} models to overcome these challenges and ensure the privacy of model parameters. The primary goal is to reduce the level of trust placed on aggregate cloud services, preserving data privacy while ensuring the correctness of computations. These approaches enable clients to train \ac{FL} models in a highly distributed fashion, and verify the aggregation process. Moreover, these trusted layers improve the integrity and privacy of the \ac{FL} computation, ensuring that clients' data remains unrevealed during the training. Additionally, verification mechanisms can help identify potential malicious behavior or tampering attempts.

Recent works in this area adopted approaches based on commitments and homomorphic hash functions~\cite{xu2019verifynet, han2021verifiable}, enabling the verification of the correct aggregation of client models performed by the aggregator~\cite{zhang2022towards}. This verification process prevents the aggregator from engaging in malicious activities such as model manipulation and weight tampering. However, as a consequence, it introduces an overhead due to the additional verification time. Current research explores new methods to verify the computation while minimizing overhead time.

One of the most comprehensive works in this area is RoFL~\cite{rofl}. This secure \ac{FL} framework enables secure aggregation via ElGamal commitments and input validation through \ac{ZK} proofs. Clients encode their updates using ElGamal commitments, which offer computational hiding, binding, and additive homomorphism under the Discrete Logarithm assumption. RoFL uses \ac{ZK} proofs to validate inputs, which allows the server to verify clients' updates without seeing their private data. 
While this method provides strong security, it also introduces limitations. For example, \ac{ZK} proofs require considerable computational power, which can make scalability challenging. Additionally, the server decodes the aggregated updates by solving a discrete logarithm, which is only feasible due to the constrained aggregation space (i.e., 32-bit integers). However, the authors do not specify how this decoding step is efficiently performed—since solving discrete logarithms at scale should be computationally expensive in practice, even within a small result space.

Another relevant work is LightVeriFL by Buyukates et al.~\cite{buyukates2022lightverifl}, which proposes a secure and verifiable aggregation protocol allowing clients to verify the aggregator's results using linearly homomorphic hashes and Pedersen commitments. Clients first mask their model updates, then compute hashes and commit them. Before sending the commitments to the server, clients must share them with each other. This step increases communication overhead and requires more complex coordination, leading to a worsening of scalability and efficiency.

There is a notable lack of research on finding a comprehensive solution that addresses attacks by the aggregator in the \ac{FL} setting while enabling clients to verify the correctness of protocol computations.
While recent works have proposed various cryptographic mechanisms to ensure integrity, such as \ac{ZK} proofs, homomorphic hashes, or commitment schemes, these approaches often introduce more complexity or communication overhead. In contrast, our approach aims for simplicity and practicality. By building on standard secure aggregation and integrating a lightweight authenticator scheme, we ensure verifiability without compromising efficiency or scalability, making the solution more accessible for deployment in real-world systems.
\section{Preliminaries}
In this section, we briefly present the basic primitives
used in our construction. These are the Secure Aggregation and the Context-Hiding Identity-based Multi-Key Linearly Homomorphic Authenticators schemes.

\label{sec:preliminaries}
\subsection{Secure Aggregation}
\label{subsec:secure_aggregation}
Several techniques have been developed to guarantee input privacy in \ac{FL}, protecting data while preserving model accuracy. Our construction relies on one of these techniques to protect the privacy of the model's input while guaranteeing the verifiability and integrity of the result. The techniques that can be used in our construction are, but are not limited to, Differential Privacy-based aggregation techniques and masking-based aggregation techniques.

Differential Privacy-based aggregation techniques~\cite{lu_dpfl, zhao_ldpfl} are based on a data perturbation method that can be realized by adding noise to the FL training parameters, according to statistical data distribution mechanisms before sharing them with the central server. This ensures that individual contributions cannot be easily reconstructed from the aggregated updates, providing provable privacy guarantees.

Masking-based aggregation techniques~\cite{Bonawitz_psa_data, Bonawitz_psa} can be adapted to obfuscate data to prevent the server from accessing raw data or model information directly and fully preserve data privacy. Hence, clients can mask their model updates and then upload them to the central server for the secure aggregation phase. Therefore, the aggregator merges the received masked model, and thanks to the specific way in which masks are generated, the obfuscations are automatically canceled in the aggregated result. This technique prevents the server from learning a single client's contributions, reducing the risk of data leakage while still enabling collaborative training.

In our implementation, to achieve privacy we used a masking-based secure aggregation protocol due to its efficiency and ability to protect individual model updates without requiring heavy cryptographic operations. Specifically, we adopt a modified version of the protocol introduced by Bonawitz et al.~\cite{Bonawitz_psa}, which enables the server to compute the aggregate model update without learning any individual client's contribution.

This protocol consists of the following algorithms:
\begin{itemize}
    \item $KeyGen()$: Each client $u$ generates pairwise shared secrets with every other client $v$ in the system using a secure key exchange protocol (e.g., Diffie-Hellman key exchange). The shared secret between clients $u$ and $v$ is denoted as $z_{u,v} = z_{v,u}$. These pairwise keys enable clients to generate correlated pseudorandom masks that will algebraically cancel during aggregation.
    \item $Mask(x_u, \{z_{u,v}\})$: Given client $u$'s local model update $x_u$ and the set of pairwise shared secrets $\{z_{u,v}\}$, this algorithm computes a masked version of the update. Client $u$ generates its pseudorandom mask as $r_u = \sum_{v \in S, v > i} PRG(k_{u,v}) - \sum_{v \in S, v < u} PRG(k_{u,v})$, where $S$ is the set of active clients in the round and $PRG$ is a pseudorandom function. The masked update is computed as $\tilde{x}_u = x_u + r_u$. This construction ensures that the pseudorandom components will sum to zero across all clients while individual updates remain hidden from the server.
    \item $Unmask(\{\tilde{x}_u\}_{u \in S})$: This algorithm is executed by the server after collecting all masked updates from the client set $S$. The server computes the aggregate as $\sum_{u \in S} \tilde{x}_u = \sum_{u \in S} (x_u + r_u) = \sum_{u \in S} x_u + \sum_{u \in S} r_u$. By the construction of the pairwise masks, $\sum_{u \in S} r_u = 0$ since each pseudorandom value $PRF(k_{u,v})$ appears exactly twice in the sum with opposite signs (once positive for the client with smaller index, once negative for the client with larger index). Therefore, the server recovers the true aggregate $\sum_{u \in S} x_u$ without learning any individual client contribution.
\end{itemize}

The methods presented here are described in a simplified way to illustrate their role in our protocol’s construction; for a more detailed explanation, please refer to~\cite{Bonawitz_psa}.

\subsection{Homomorphic Authenticators}
\label{subsec:hom_auth}

Multi-Key Homomorphic Authenticators schemes enable multiple parties to sign data and homomorphically aggregate them, keeping individual contributions private. Therefore, signatures generated by different parties can be combined into a final aggregated signature that can be verified using the combined data without revealing the original individual signatures. This property is particularly useful in distributed computing scenarios where multiple parties contribute to a computation while preserving data confidentiality. In the following sections, we use “sign” and “authenticate” interchangeably.

Our protocol adopts the Context-Hiding Identity-based Multi-Key Linearly Homomorphic Authenticators scheme presented by Schabhüser et al.~\cite{schabh}. The scheme provides the following properties that make it suitable for our construction:

\subsubsection{Multi-Key Signatures} Each participant owns a private key derived from their identity and can independently generate a partial authenticator for a message. Therefore, these authenticators, generated by different parties, can later be combined into a single, cryptographically valid signature representing the aggregated computation result. This property removes the need for a shared signing key or trusted setup among participants.

\subsubsection{Linearly Homomorphic} The scheme supports linear operations directly on authenticated data, meaning that if individual signatures $\sigma_1, \sigma_2, \ldots, \sigma_n$ are signatures of the authenticated messages $m_1, m_2, \ldots, m_n$ respectively, then the combined signature $\sigma_{agg}$ obtained through the evaluation algorithm will authenticate the linear combination $m_1 + m_2 + \cdots + m_n$.

\subsubsection{Context-Hiding} This property provides some measure of input privacy; once signed, individual contributions remain hidden, and no one can leak information from signatures. Hence, the input values are private concerning some outside party (external entity), and an identity that provides inputs to the computation (internal entity).

This protocol consists of the following algorithms:
\begin{itemize}
\item $Setup(1^\lambda$): On input of a security parameter $\lambda$, this algorithm generates the system's public parameters $pp$. These parameters are published and used by all participants in the system.

\item $KeyGen(pp, id)$: On input the public parameters $pp$ and a unique identity $id \in \mathbb{N}$, this algorithm generates the identity-specific secret key $sk_{id}$ and the corresponding verification key $vk_{id}$.

\item $Auth(sk_{id}, x_u)$: On input a secret key $sk_{id}$ and a client $u$'s vector of local model update $x_u \in \mathbb{R}^d$, where $d$ is the length of the vector, this algorithm outputs an authenticator $\sigma_u = (\Lambda, R, S)$. Here, $\Lambda$ contains identity-specific components that bind the signature to the signer's identity, while $R$ and $S$ are global components that enable homomorphic operations across different signers' authenticators.

\item $Eval(\{\sigma_u\}_{u \in S})$: On input a set of authenticators $\{\sigma_u\}_{u \in S}$, where $S$ is the set of active clients in the round, this algorithm computes the homomorphic evaluation $\sigma_{agg} = \sum_{u \in S} \sigma_u$ that authenticates the linear combination $x_{agg} = \sum_{u \in S} x_u$ without access to the individual vector of model updates $x_u$. The method preserves the context-hiding property.

\item $Ver(\{vk_{id}\}_{id \in S}, x_{agg}, \sigma_{agg})$: On input a set of verification keys corresponding to the identities involved in the computation, an aggregated vector of model updates $x_{agg}$, and an authenticator $\sigma_{agg}$, this algorithm outputs $1$ if and only if $\sigma_{agg}$ is a valid aggregated signature for the vector of aggregated model updates $x_{agg}$ under the specified set of identities. These identities are the contributors of the aggregated authenticator $\sigma_{agg}$. The verification process confirms both the authenticity of individual contributions and the correctness of the homomorphic aggregation operation.
\end{itemize}

The presented methods have been described in a simplified way to introduce their usage in the construction of our protocol; we refer to~\cite{schabh} for a more complete description.
\section{Verifiable Federated Learning}
\label{sec:protocol}

In this section, we present our construction, a protocol designed to enable clients to securely train a global \ac{FL} model while ensuring the integrity and correctness of the aggregated model parameters. Specifically, our approach allows clients to verify the correctness and integrity of the aggregated model parameters while relying on an underlying secure aggregation scheme to preserve data confidentiality. To achieve data integrity, we leverage the Context-Hiding Identity-based Multi-Key Linearly Homomorphic Authenticators scheme (detailed in Section~\ref{subsec:hom_auth}) that allows clients to validate in an independent way the aggregated results. If the aggregator or other potential adversaries attempt to interfere with or manipulate the training of the \ac{FL} global model, our protocol is able to detect any unauthorized modifications.

In the following, we define the protocol in detail, starting with an overview of the involved parties and how they interact. We then outline the different phases of the protocol and the data that are exchanged by the parties at each phase.

\subsection{Construction}
\label{subsec:protocol}

\begin{algorithm}[tb]
    \caption{Secure and Verifiable Federated Aggregation Protocol}
    \begin{algorithmic}[1]
    
    \Statex \textbf{Phase 0: Initialization} \\
        Each client:
            \begin{itemize}
                \item Generates a key pair $(sk_{id}, vk_{id})$ for signing.
                \item Publishes the public verification key $vk_{id}$ along with its identifier.
                \item Agrees on protocol parameters for secure masking.
            \end{itemize}

    \Statex \textbf{Phase 1: Client Preparation} \\
        Each client $k$:
            \begin{itemize}
                \item Trains its local model on private data obtaining $X^{(k)} \in \mathbb{R}^{d \cdot m}$.
                \item Masks each element of the model update matrix:
                \begin{equation}
                    c_{i,j}^{(k)} = \text{Mask}(x_{i,j}^{(k)}, \{z_{u,v}\})
                \end{equation}
                \item Computes an authenticator for each column:
                \begin{equation}
                    \sigma_j^{(k)} = \text{Auth}(sk_{id}, x_j^{(k)})
                \end{equation}
                \item Sends to the aggregator:
                \begin{itemize}
                    \item Masked model updates $\{c_{i,j}^{(k)}\}$.
                    \item Column authenticators $\{\sigma_j^{(k)}\}$.
                \end{itemize}
            \end{itemize}

    \Statex \textbf{Phase 2: Aggregation} \\
        The aggregator:
            \begin{itemize}
                \item Aggregates the masked model updates from all active clients $S$:
                \begin{equation}
                    x_{i,j} = \text{UnMask}(\{c_{i,j}^{(k)}\}_{k \in S})
                \end{equation}
                \item Aggregates the authenticators:
                \begin{equation}
                    \sigma_j = \text{Eval}(\{\sigma_j^{(k)}\}_{k \in S})
                \end{equation}
                \item Sends the aggregated model update vectors $\{x_j\}$ and corresponding authenticators $\{\sigma_j\}$ to all clients.
            \end{itemize}

    \Statex \textbf{Phase 3: Client Verification} \\
        Each client:
            \begin{itemize}
                \item Verifies the integrity of the aggregated updates:
                \begin{equation}
                    check_j = \text{Ver}(\{vk_{id}\}_{id \in S}, x_j, \sigma_j)
                \end{equation}
                \item If all verifications succeed:
                \begin{itemize}
                    \item Accepts the aggregated result.
                    \item Computes the average: $x_{i,j} = x_{i,j} / |S|$.
                    \item Proceeds to the next round.
                \end{itemize}
                \item Otherwise:
                \begin{itemize}
                    \item Rejects the result due to possible aggregator misbehavior.
                \end{itemize}
            \end{itemize}
    \end{algorithmic}
    \label{algo:phases}
\end{algorithm}

\begin{figure*}[tb]
    \centering
    \includegraphics[width=\textwidth]{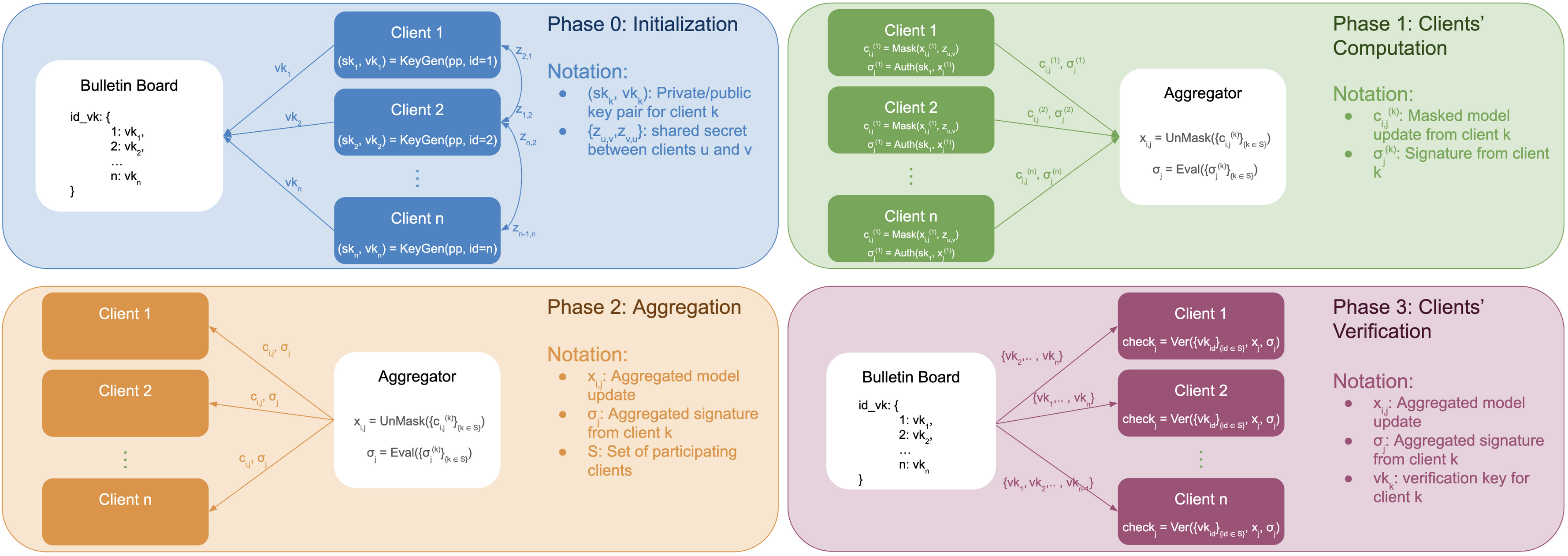}
    \caption{Overview of the protocol phases. In the Initialization phase, clients agree on shared secrets, generate key pairs, and publish verification keys to the Bulletin Board. During the Clients' Computation phase, each client masks and computes authenticators over its local model update. In the Aggregation phase, the server aggregates the masked updates and the corresponding authenticators. Finally, in the Client's Verification phase, each client verifies the correctness of the aggregated result using the returned authenticators. For simplicity, the illustration shows a single protocol round, assuming that all participants follow the same procedure.}
    \label{fig:phases}
\end{figure*}

Our protocol requires an initialization phase conducted by a trusted entity. During this phase, the necessary cryptographic parameters are generated and made publicly available to all the involved parties. We assume that the trusted entity may safely go offline once this phase is complete.

Using the published public parameters, each client can independently generate a unique identity $id$ and a key-pair consisting of a secret key and the corresponding verification key, as follows:

\begin{equation}
    (sk_{id}, vk_{id}) = KeyGen(pp, id)
\end{equation}

Additionally, we assume the existence of a public key infrastructure that allows clients to register their identities. Specifically, each client publishes its identity $id$ and the corresponding verification key $vk_{id}$ to a publicly accessible bulletin board~\cite{Bonawitz_psa, EIFFeL}. This mechanism prevents any adversary, such as the aggregator or external entities, from impersonating honest clients. To avoid the usage of a third-party public bulletin board and to reduce the risks associated with their potential compromise, solutions based on blockchain can be used as an alternative~\cite{bulletin_blockchain}.

The system involves the active collaboration of multiple parties, as shown in Figure~\ref{fig:phases}. Two of them concretely contribute to the computation of the global FL model: the aggregator and a variable number of clients.
The clients are responsible for training the global \ac{FL} model. They compute the model parameters locally, authenticate, and mask them. Once done, they request the aggregator to perform the necessary operations to obtain the final global \ac{FL} model. The clients, optionally, could verify the correctness of the resulting model to ensure that the protocol has been followed accurately and without tampering.
The aggregator's primary role is to perform the aggregation of the clients' weights and authenticators, and then return the results of its computation back to the clients. The aggregator can be honest but curious, attempting to infer information from the client's parameters or acting as a malicious entity, trying to undermine the protocol's integrity.

Our protocol is structured into different phases, outlined in Algorithm~\ref{algo:phases}. These phases are executed equally in each round, with the exception of the initialization phase, which is performed prior to the protocol's beginning. To simplify, we describe the protocol's execution within a single round, assuming that all participating parties perform identical operations during each iteration.

The initialization phase, as previously mentioned, takes place before the protocol execution and is performed only once.

Each round of the protocol starts with the Training Phase, during which each client $k$ trains its local \ac{FL} model on private data and obtains a matrix of model updates $X^{(k)} \in \mathbb{R}^{d \cdot m}$, where $d$ is the number of rows and $m$ is the number of columns. It is assumed that each client generates a fixed number of model updates in every round.

To maintain simplicity in notation and clarity in protocol flow, we adopt the following notational conventions. We use $x_{ij}^{(k)} \in \mathbb{R}^{d \cdot m}$ to denote a single value, corresponding to the $i-th$ element of the $j-th$ column of the model update matrix $X^{(k)} \in \mathbb{R}^{d \cdot m}$. Similarly, we use $x_j^{(k)} \in \mathbb{R}^d$ to denote the $j-th$ column of $X^{(k)} \in \mathbb{R}^{d \cdot m}$. This notation will remain consistent throughout the rest of the description.

In the Client Preparation Phase, each client prepares its local model updates for secure and verifiable aggregation. This phase ensures both the privacy of individual model updates and the integrity of the global aggregation.

Each client performs two key operations on its model update matrix $X^{(k)} \in \mathbb{R}^{d \cdot m}$. First, to protect the confidentiality of individual updates, the client masks each element of the matrix using a masking-based Secure Aggregation scheme, as described in Section~\ref{subsec:secure_aggregation}. Specifically, for each element $x_{i,j}^{(k)}$ the client computes:

\begin{equation}
    c_{i,j}^{(k)} = \text{Mask}(x_{i,j}^{(k)}, \{z_{u,v}\})
\end{equation}

where $\{z_{u,v}\}$ denotes the shared random values used in the masking procedure. This ensures that the server learns only the aggregated result and cannot infer any individual client's data.

Second, to enable post-aggregation verification, the client generates an authenticator for each column $x_{j}^{(k)} \in \mathbb{R}^d$ of its model update matrix using the homomorphic signature scheme presented in Section~\ref{subsec:hom_auth}. The authenticator is computed as:

\begin{equation}
    \sigma_j^{(k)} = Auth(sk_{id}, x_{j}^{(k)})
\end{equation}

where $sk_{id}$ is the private key of the client. This authenticator allows any verifier to later check that the client's contribution has been correctly included in the final aggregate. Clients authenticate vectors rather than individual values in order to minimize communication and computational overhead. This optimization increases efficiency while maintaining security guarantees.

Finally, each client transmits both the masked model update vectors and their corresponding authenticators to the aggregator for secure aggregation and subsequent verification.

The second phase of the protocol is the Aggregation Phase, during which the aggregator processes the masked model updates and authenticators received from all participating clients in order to compute the global model updates. Specifically, the aggregator performs the following operations:
\begin{equation}
    x_{i,j} = UnMask(\{c_{i,j}^{(k)}\}_{k \in S})
\end{equation}
\begin{equation}
    \sigma_j = Eval(\{\sigma_j^{(k)}\}_{k \in S})
\end{equation}
where $S$ denotes the set of active clients in the current round. The first operation reconstructs the aggregated model update for the $(i,j)-th$ entry by removing the masking applied during the Secure Aggregation phase. The second operation aggregates the individual authenticators $\sigma_j^{(k)}$ provided by the clients into a single authenticator $\sigma_j$, using the homomorphic properties of the underlying signature scheme.

The final phase of the protocol is the Client Verification Phase, in which each client (optionally) verifies the correctness of the aggregation performed by the aggregator. The aggregator broadcasts the aggregated model updates and the corresponding aggregate authenticator to all clients. Upon receiving this information, each client performs a verification step to ensure that the aggregation has been carried out honestly and without manipulation. Each client computes the following:

\begin{equation}
    check_j = Ver(\{vk_{id}\}_{id \in S}, x_{j}, \sigma_j)
\end{equation}

where $\{vk_{id}\}_{id \in S}$ are the public verification keys of the participating clients in the set $S$, $x_j$ is the aggregated update vector for column $j$, and $\sigma_j$ is the corresponding aggregate authenticator. The function $Ver$ denotes the verification algorithm associated with the employed homomorphic signature scheme.

If the verification succeeds, the client accepts the result and proceeds to compute the final average by dividing each aggregated value by the number of participating clients:

\begin{equation}
    x_{i,j} = x_{i,j} / |S|
\end{equation}

Subsequently, the client updates its model and proceeds to the next training round. However, if the verification fails for any column, it indicates potential misbehavior by the aggregator, such as incorrect or malicious computation. In such cases, the client rejects the aggregation result, and the round is considered invalid.

\subsection{Security Consideration}
\label{subsec:security_consideration}
Our protocol addresses two key properties in the context of \ac{FL}: the privacy of the client updates and the verifiability of the aggregation process. The first aspect ensures that individual model updates remain confidential and are guaranteed by the underlying secure aggregation scheme, which prevents the server or any unauthorized party from accessing raw client data. The second aspect guarantees the integrity of the aggregation process and is achieved through the use of the Context-Hiding Identity-based Multi-Key Linearly Homomorphic Authenticators scheme, which allows clients to verify the correctness of the aggregation performed by the server.

In the following, we analyze the security guarantees provided by our construction under two adversarial models: a semi-honest aggregator, which follows the protocol but attempts to infer private information from the clients, and a malicious aggregator, which may deviate arbitrarily from the protocol.

\subsubsection{Semi-honest aggregator}
In the semi-honest setting, the aggregator behaves according to the protocol but may attempt to infer private information from the data it receives. Our protocol protects against such privacy threats through the secure aggregation mechanism, which includes a masking phase that obfuscates each client's model updates before transmission. As a result, the aggregator operates only on masked data and cannot infer any information about the original client updates, even through passive analysis or inference attacks. The confidentiality of individual contributions is preserved throughout the aggregation process.

\subsubsection{Malicious aggregator}
In the malicious setting, the aggregator may attempt not only to learn sensitive data, but also to actively tamper with the protocol. This includes modifying or injecting updates, manipulating the final model, or incorrectly aggregating values and signatures.

Our protocol mitigates these threats by enabling each client to verify the correctness of the aggregated result using the Context-Hiding Identity-based Multi-Key Linearly Homomorphic Authenticators scheme. Before transmitting model updates, each client authenticates its model updates using its private key. The aggregator performs aggregation directly over the masked values and the authenticators. Due to the homomorphic property of the signature scheme, any unauthorized modification — such as altering the aggregated vector $x_{agg}$ or injecting a malicious value $e$ (e.g., setting $x_{agg}' = x_{agg} + e$ or $x_{agg}' = e$) — will cause the resulting signature to be invalid. Clients will detect such manipulation during the verification phase.

Moreover, attempts to improperly aggregate authenticators are also detectable. For instance, if the aggregator introduces a forged signature $\sigma_e$ for an injected value $e$, and computes a new aggregated signature $\sigma_{agg}' = Eval(\{\sigma_j^{(k)}\}_{k \in S} \cup {\sigma_e})$, clients will reject it. This is because every client verifies the final result using the public verification keys $\{vk_{id}\}_{id \in S}$ that were published to the publicly accessible bulletin
board (or to a blockchain) before aggregation. Any deviation from the valid set of signing keys results in a verification failure.

This construction ensures that the aggregator cannot forge valid signatures on unauthorized updates without access to private keys. As a consequence, clients can verify if the returned model has been honestly computed. If verification fails, the protocol halts, and the compromised round is discarded.

For a formal security proof of the masking-based Secure Aggregation scheme's properties, including input confidentiality and robustness against collusion, we refer to~\cite{Bonawitz_psa}. For a formal security proof of the Context-Hiding Identity-based Multi-Key Linearly Homomorphic Authenticators scheme's properties, including unforgeability and context hiding, we refer to~\cite{schabh}.
\section{Experimental Evaluation}
\label{sec:experiments}

We now present the implementation details of our construction and the experimental configuration used to evaluate its performance. We then describe the evaluation methodology and report results obtained by testing our protocol with a few million model updates.

\subsubsection{Experimental Setup}
Our protocol is implemented in the Rust programming language, utilizing the Arkworks library~\cite{arkworks} for efficient cryptographic operations. Specifically, we employ the BLS12-381 pairing-friendly elliptic curve through the ark-bls12-381 module. All experiments were conducted on a laboratory virtual machine equipped with 32 vCPUs and 128 GB of RAM, powered by 2 CPU Intel Xeon 4316 2.3 GHz.


\subsection{Federated Learning Accuracy}
The common requirement in cryptography that floating point numbers must be truncated to a fixed precision for later encoding into an integer format results in a twofold influence on the protocol: the lower the handled precision results in faster commitment computation but can cause regular FL accuracy to drop.
We analyze the effect on performance on three datasets: MNIST, CIFAR, and Merged Federated MNIST from the LEAF benchmark~\cite{caldas2018leaf}.
The Merged Federated MNIST corresponds to the original LEAF benchmark dataset while merging classes corresponding to upper and lower case characters with semantically identical content.
Following the original protocol in~\cite{cohen2017emnist}, classes C, I, J, K, L, M, O, P, S, U, V, W, X, Y, and Z are merged with their lowercase versions.

We can see the results in Table \ref{tab:dec-precision} with four decimal places of precision being sufficient for identical performance to running FL without the protocol and hence maintaining full floating point precision. This level of precision is the one we use in the following cryptographic time evaluation experiments.

\begin{table*}[tb]
    \centering
    \begin{tabular}{ c c c c c c}
    \toprule
      \textbf{Dataset} & \makecell{ \textbf{Regular} \\ \textbf{Accuracy}} & \multicolumn{4}{c}{\textbf{Protocol Accuracy}} \\
      \cmidrule{3-6} & & 2 DP & 4 DP   &  6 DP & 8 DP\\
      \cmidrule{3-6}
      MNIST &  $97.49 \pm 0.16$  & $91.36 \pm 0.56$ & $97.49 \pm 0.10$ &  $97.60 \pm 0.04$ & $97.40 \pm 0.10$  \\
      CIFAR &  $79.13 \pm 0.45$  & $10.44 \pm 0.64$  & $79.55 \pm 0.76$ &  $79.10 \pm 0.85$ & $79.41 \pm 0.78$  \\
      \makecell{Merged \\ FEMNIST}     &  $88.17 \pm 0.52$  & $43.70 \pm 19.21$ & $88.21 \pm 0.51$ &  $88.16 \pm 0.36$ & $87.95 \pm 0.28$  \\
    \bottomrule
    \addlinespace[0.3cm]
    \end{tabular}
    \caption{Accuracy comparison for different levels of rounding to a given number of decimal places (DP). Results are averaged over $5$ runs with associated standard deviation. Higher precision results in slightly increased computation time for authenticators.}
    \label{tab:dec-precision}
\end{table*}

\subsection{Cryptographic Time Evaluation}
Significant computational overhead may be introduced when performing cryptographic operations over large-scale neural network models. To evaluate the computational efficiency of our verification protocol, we measure the execution time of key cryptographic operations introduced by our construction, focusing on the linearly homomorphic authenticator scheme described in Section~\ref{subsec:hom_auth}.  Specifically, we consider (i) authenticator generation by clients during the Client's Computation phase, (ii) aggregation of authenticators by the server in the Aggregation phase, and (iii) verification of the aggregated authenticator by clients in the Client's Verification phase.

We focus on these components because they are the primary parts of our verifiability mechanism, and their computational cost impacts the real-world practicality of the proposed solution. Since the underlying secure aggregation scheme used to protect client privacy is a well-established and extensively studied method, and its performance has been analyzed in prior works,
and we do not include its benchmarks here. Consequently, our results highlight the overhead introduced specifically by the authenticator scheme on top of existing secure aggregation to ensure integrity and verifiability within the \ac{FL} framework.

\subsubsection{Client's Computation}
\begin{figure}[tb]
    \centering
    \includegraphics[width=\columnwidth]{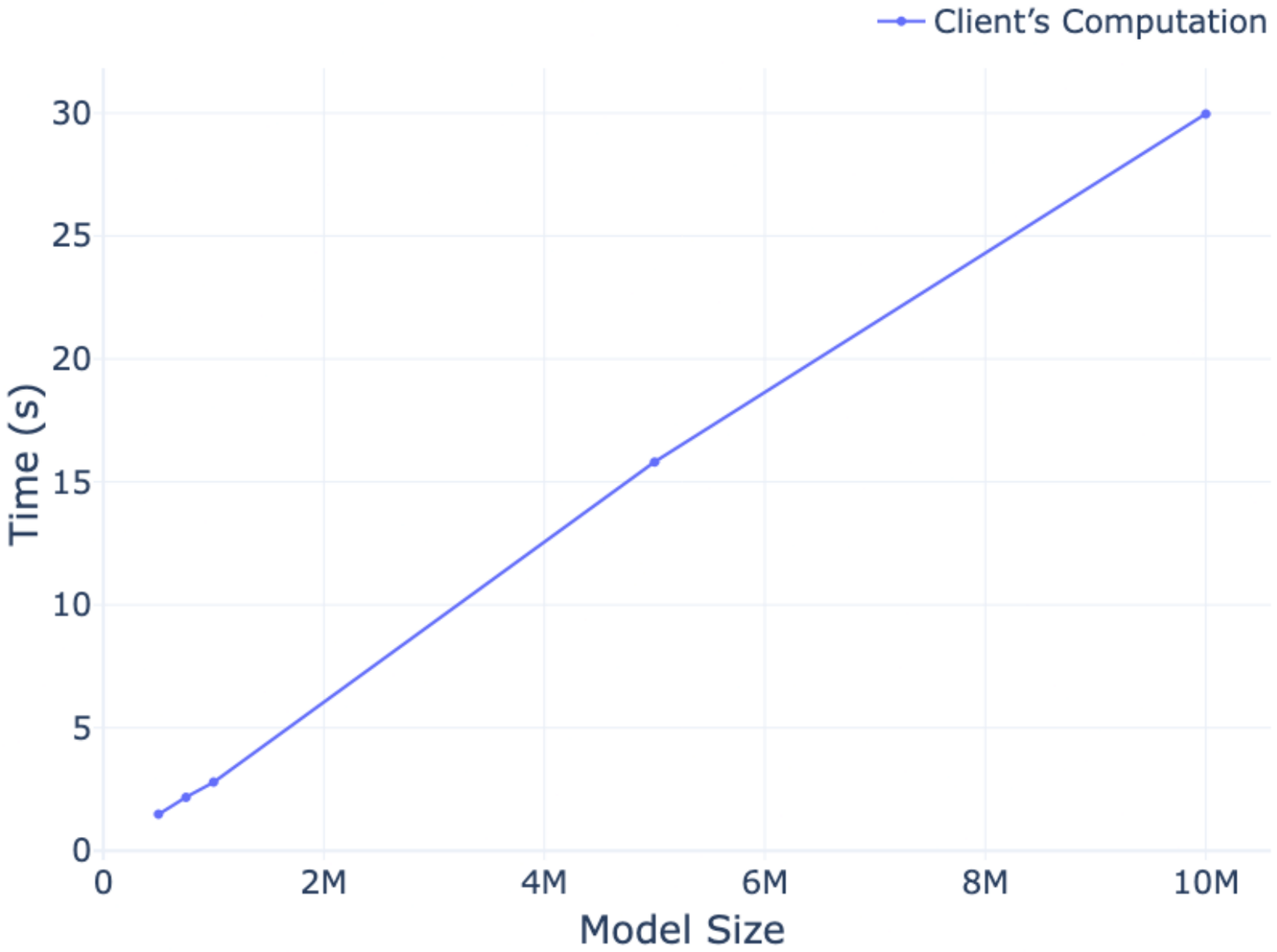}
    \caption{Computation overhead in the Clients' Computation phase. The plot reports the average time a single client requires to generate authenticators over model updates of varying sizes. Results are averaged over 10 runs.}
    \label{fig:computation}
\end{figure}

In this phase, each client generates an authenticator over its model parameters using the Context-Hiding Identity-based Multi-Key Linearly Homomorphic Authenticator scheme before sending them to the aggregator. This process ensures that clients can later verify the integrity of the aggregated result without revealing their private model updates.
This step introduces additional computational overhead compared to standard \ac{FL} but remains efficient in practice. Our experiments show that generating authenticators for a model with one million parameters takes, on average, 2.8 seconds per client (see Figure~\ref{fig:computation}). 

The authenticator scheme operates on vectors rather than individual values, allowing clients to batch model parameters and reduce the number of cryptographic operations. This design also improves parallelism: each column of the model update can be split into multiple sub-columns (agreed upon by all clients) and processed independently. As a result, clients can better exploit multi-core architectures to compute authenticators concurrently. The number of columns can be tuned based on system capabilities (i.e., the number of available CPU cores), improving overall performance.

\subsubsection{Aggregation}
After clients submit their masked model updates and corresponding authenticators, the aggregator performs aggregation on both. Thanks to the homomorphic properties of the authenticator scheme, individual authenticators can be combined into a single aggregated authenticator corresponding to the aggregate of the clients' updates.

\begin{figure}[tb]
    \centering
    \includegraphics[width=\columnwidth]{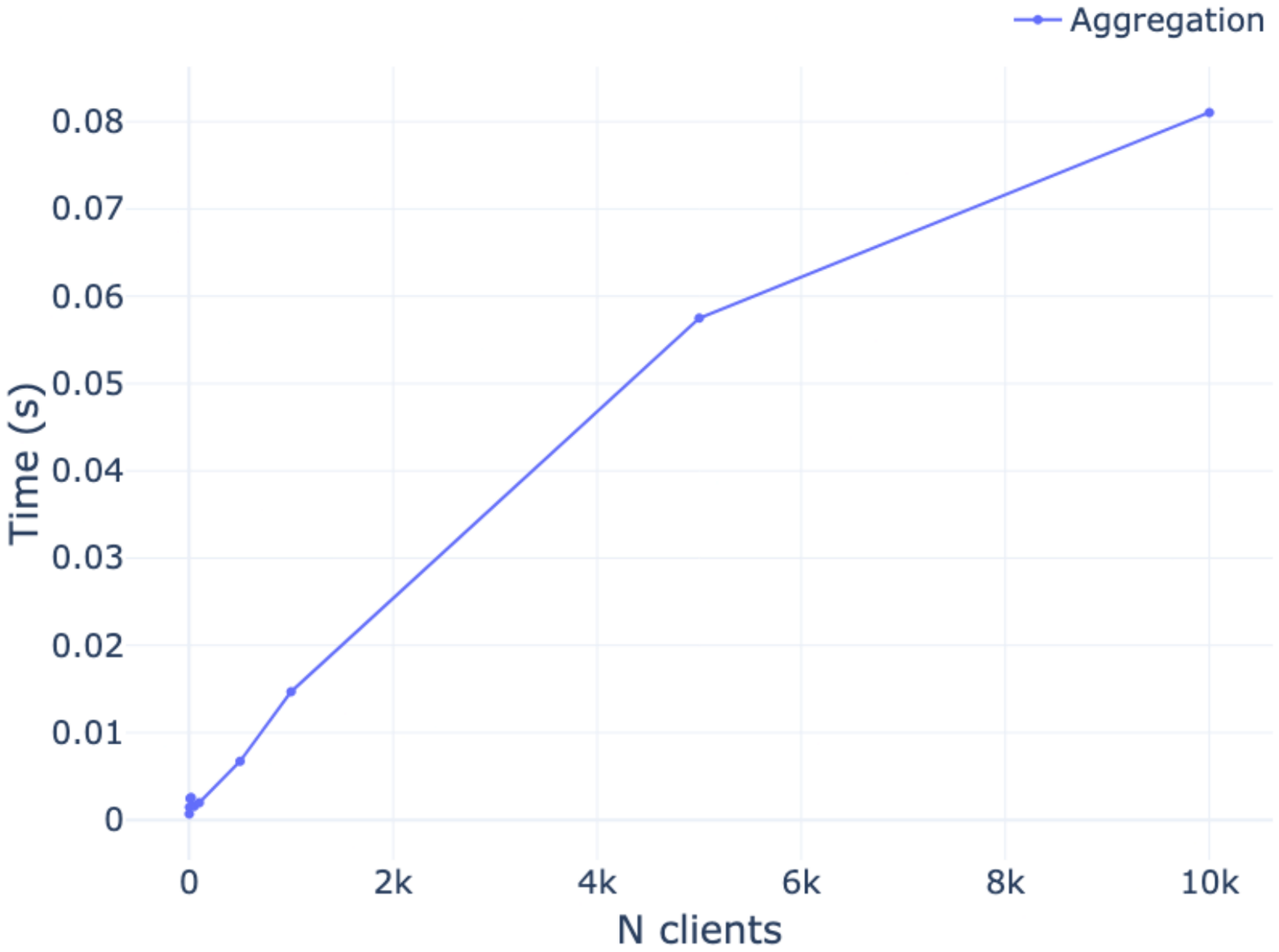}
    \caption{Computation overhead in the Aggregation phase. The plot reports the average time required by the server to aggregate authenticators from different numbers of clients. Results are averaged over 10 runs.}
    \label{fig:aggregation}
\end{figure}

Our experimental results show that the computational cost of this aggregation phase scales with the number of participating clients in a \ac{FL} round. Specifically, it takes approximately 2 milliseconds to aggregate authenticators for 100 clients, around 6 milliseconds for 500 clients, and about 15 milliseconds for 1000 clients (see Figure~\ref{fig:aggregation}). These results show that our protocol remains computationally efficient even as the number of clients increases.

\subsubsection{Client's Verification}
After receiving the aggregated model update and its corresponding aggregated authenticator, each client can perform the verification to ensure the correctness of the aggregator's computation. This verification is essential to detect any malicious manipulation by the aggregator.

\begin{figure}[tb]
    \centering
    \includegraphics[width=\columnwidth]{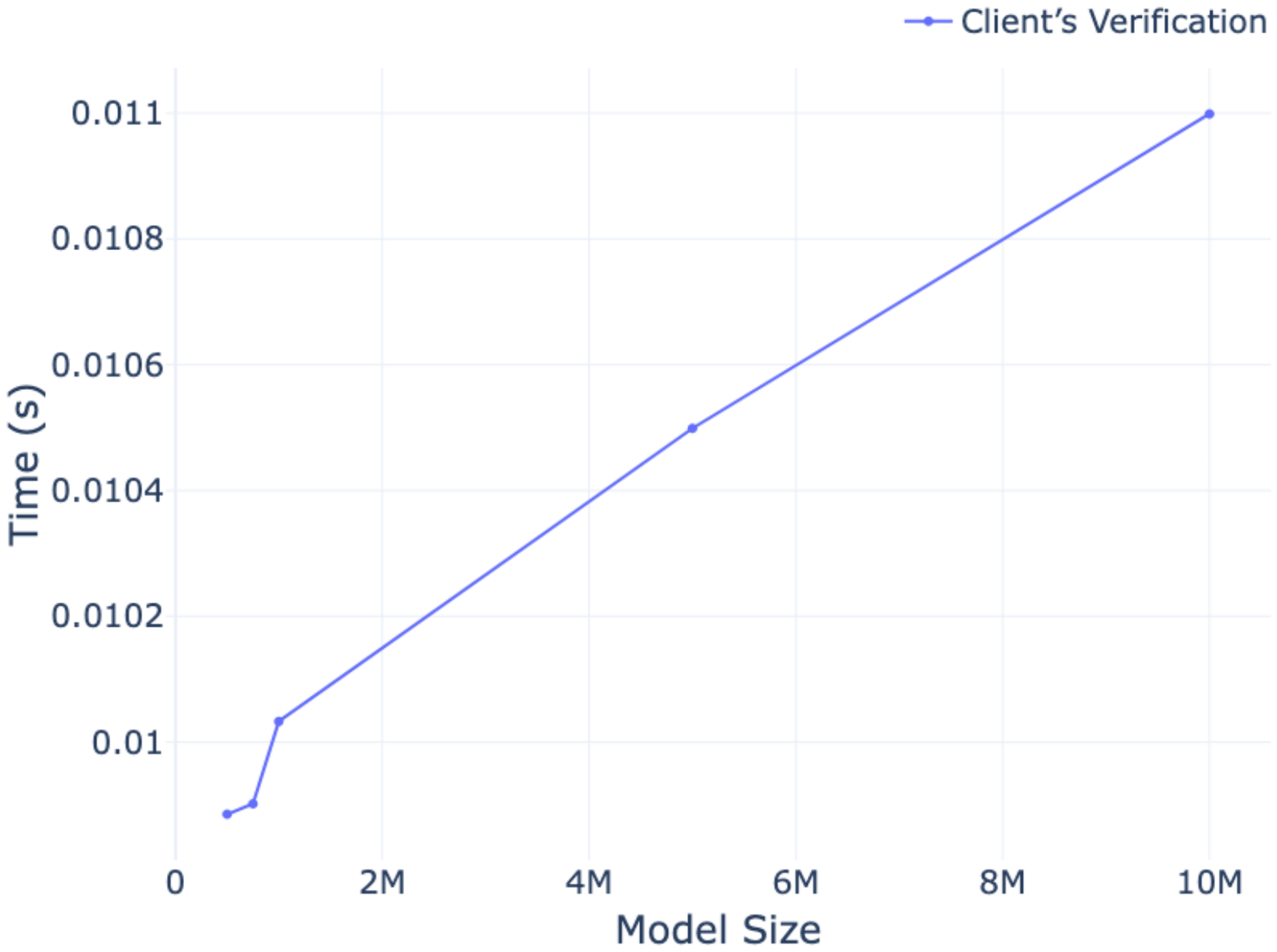}
    \caption{Computation overhead in the Client’s Verification phase. The plot shows the average time required by a single client to verify the aggregated authenticator for model updates of varying sizes. Results are averaged over 10 runs.}
    \label{fig:verification}
\end{figure}

Our results show that, on average, verifying the aggregated authenticator for a model with one million parameters requires approximately 10 milliseconds for a single client (see Figure~\ref{fig:verification}).

\subsection{Communication Overhead}
Our protocol introduces communication overhead compared to standard \ac{FL}. Each client sends its masked model update along with a set of authenticators (one per column of the model update). The server returns the aggregated model update and the corresponding aggregated authenticators. Since the number of columns is usually small, the additional communication cost remains small.

\subsubsection{Client-to-Server Communication}
Each client sends authenticators corresponding to their model updates to the server. The update vector is composed of $m$ columns of size $d$, resulting in each client transmitting $m$ authenticators — one per column. The size of each authenticator is constant and does not depend on the column length $d$, thus the overall communication overhead grows linearly with the number of columns $m$. As shown in Table~\ref{tab:comm_complexity}, a client sending authenticators for a model with 1 million parameters results in approximately 0.28 KB per authenticator transmitted. 

\subsubsection{Server-to-Client Communication}
The server returns the aggregated masked model update, which has the same size as a single client's update, along with a single aggregated authenticator. This minimizes the communication overhead since the transmitted data does not scale with the number of participants.

\begin{table}[ht]
    \centering
    \begin{tabular}{c c c c}
        \toprule
        \textbf{Model Size} & \makecell{\textbf{N. Columns} \\ \textbf{($m$)}} & \makecell{\textbf{Column Length} \\ \textbf{($d$)}} & \makecell{\textbf{Authenticator Size} \\ \textbf{per column} (KB)} \\
        \midrule
        500,000 & 10 & 50,000 & 0.28 \\
        750,000 & 10 & 75,000 & 0.28 \\
        1,000,000 & 10 & 100,000 & 0.28 \\
        5,000,000 & 10 & 500,000 & 0.28 \\
        10,000,000 & 10 & 1,000,000 & 0.28 \\
        \bottomrule
        \addlinespace[0.3cm]
    \end{tabular}
    \caption{Communication overhead for varying model sizes. Both clients and the aggregator send one authenticator per column.}
    \label{tab:comm_complexity}
\end{table}
\section{Conclusion}
\label{sec:conclusion}
We have introduced a novel verifiable federated learning protocol that guarantees the integrity of the aggregator’s computations while preserving client privacy. Leveraging a Context-Hiding Identity-based Multi-Key Linearly Homomorphic Authenticators scheme, our protocol enables clients to efficiently verify the correctness of aggregated model updates without exposing their model update. By building on an existing secure aggregation framework, we focused on verifying aggregation integrity, providing mechanisms to detect and prevent tampering or malicious behavior by the aggregator.

Our experimental evaluation demonstrates that the overhead the authenticator scheme introduces is small. For a model with one million parameters, client authenticator generation takes approximately 2.8 seconds while the verification process takes approximately 10 milliseconds. Aggregation on the server requires around 15 milliseconds to aggregate authenticators from 1000 clients.

Furthermore, our construction can be extended to handle dropout clients, enhancing its robustness and applicability in practical \ac{FL} deployments.

\section*{Acknowledgment}
This work was supported in part by project SERICS (PE00000014) under the NRRP MUR program funded by the EU - NGEU. Views and opinions expressed are however those of the authors only and do not necessarily reflect those of the European Union or the Italian MUR. Neither the European Union nor the Italian MUR can be held responsible for them.

\bibliographystyle{IEEEtran}
\bibliography{bibliography}

\end{document}